\documentclass[12pt]{article}
\usepackage{a4}

\def \be {\begin{equation}}
\def \ee {\end{equation}}
\newcommand{\ba}{\begin{eqnarray}}
\newcommand{\ea}{\end{eqnarray}}

\def \del {\partial}
\def \dels {\partial\kern-.5em / \kern.5em}

\newcommand{\ol}{\overline}

\def \a {\alpha}
\def \b {\beta}

\def \g {\gamma}

\def \d {\delta}
\def \eps {\epsilon}

\def \lam {\lambda}

\def \rr {\raise.35ex\hbox{\small $\prime$}\kern-.17em{\mbox{\large $\imath$}}}

\def \As {{A\kern-.5em / \kern.5em}}
\def \Ds {D\kern-.7em / \kern.5em}

\newcommand{\dmu}{{\dot \mu}}
\newcommand{\dnu}{{\dot \nu}}
\newcommand{\dlam}{{\dot \lambda}}

\newcommand{\umu}{\underline{\mu}}
\newcommand{\unu}{\underline{\nu}}
\newcommand{\ulam}{\underline{\lam}}
\newcommand{\ua}{\underline{\a}}
\newcommand{\ub}{\underline{\b}}
\newcommand{\ug}{\underline{\g}}

\def \IIA {$\mbox{I\hspace{-.1em}IA\hspace{.4em}}$}

\begin{document}

\begin{titlepage}

\begin{center}

\hfill UT-08-16
\vskip .5in

\textbf{\LARGE A Concise Review on \\ 
\vskip.5cm
M5-brane in Large $C$-Field Background}

\vskip .5in
{\large
Pei-Ming Ho$^\dagger$\footnote{
e-mail address: pmho@phys.ntu.edu.tw}, 
}\\
\vskip 3mm
{\it\large
$^\dagger$
Department of Physics and Center for Theoretical Sciences, \\
National Taiwan University, Taipei 10617, Taiwan,
R.O.C.}\\
\vspace{60pt}
\end{center}
\begin{abstract}

We give a concise review of recent developments of
the M5-brane theory derived from the Bagger-Lambert model.  
It is a 6 dimensional supersymmetric self-dual gauge theory
describing an M5-brane in a large constant $C$-field background. 
The non-Abelian gauge symmetry  
corresponds to diffeomorphisms preserving 
an exact 3-form on the worldvolume, 
which defines a Nambu-Poisson bracket. 
Various interesting geometric and algebraic properties of the theory 
are discussed.

\end{abstract}

\end{titlepage}
\setcounter{footnote}{0}

\section{Introduction}

The basic building blocks of M theory are 
membranes (also called M2-branes) and M5-branes.
Membranes are fundamental objects carrying electric charges 
with respect to the 3-form $C$-field, 
and M5-branes are magnetic solitons. 
The purpose of this article is to give a concise summary 
of recent developments on a new formulation of M5-branes 
in a large constant $C$-field background. 

The bosonic field content of an M5-brane theory
should include scalars that represent M5-brane fluctuations 
in the transverse directions
and a 2-form gauge potential with self-dual field strength. 
The low energy effective action has been known \cite{oldM5} 
for some time. 
A new formulation \cite{M51,M52} was obtained from 
the BLG model \cite{BL,Gustavsson} 
by setting the Lie 3-algebra which dictates 
the gauge symmetry of the BLG model 
to be the Nambu-Poisson bracket on a 3-manifold.
\footnote{
For a review on how different choices of Lie 3-algebras 
for the BLG model lead to different physical systems, 
see \cite{HoNambu}.
}
(The 3-manifold is defined as an internal space from the M2-brane viewpoint, 
but becomes part of the M5-brane worldvolume.)
The resulting theory is a non-Abelian self-dual gauge theory 
with novel geometric and algebraic structures 
worthy of further investigations.

The description of multiple M5-branes has been a long-standing mystery. 
The new formulation of M5-brane
might shed some light on this question, 
since we expect that the multiple M5-brane theory to be 
a non-Abelian gauge theory. 
This can not be an ordinary non-Abelian gauge theory 
resembling Yang-Mills theory, 
because the entropy of coincident $N$ M5-branes 
should scale as $N^3$, 
rather than $N^2$ like Yang-Mills theories. 
Thus we expect that there is a new gauge symmetry to be discovered 
for multiple M5-branes. 
(The fact that the entropy of $N$ M2-branes scales as $N^{3/2}$ 
was recently explained in the framework of the BLG model 
in \cite{Chu:2008qv}.)
The novel gauge symmetry in the new formulation of M5-brane theory
is very likely the first step towards the discovery of 
the mysterious gauge structure for multiple M5-branes.

This article is organized as follows.
In sec. \ref{NP}, we give some motivation for why 
Nambu-Poisson structure is expected to play a role 
in M5-brane theory when there is a large $C$-field background. 
Basic algebraic properties of the Nambu-Poisson bracket 
are reviewed at the end of the section. 
The action and symmetries of the new formulation of M5-brane 
in large $C$-field background are given in sec. \ref{M5}, 
and in sec. \ref{Comments} we comment on related topics
such as Seiberg-Witten map, self-dual gauge theory, 
and connection with the old formulation of M5-brane. 
Finally, in the last subsection, we give a list of questions for 
future research.

\section{From Noncommutative space to \\ Nambu-Poisson structure}
\label{NP}

It is well-known that D-branes in constant $B$-field background 
are to be described as noncommutative gauge theories 
\cite{ChuHo,Schomerus,SeibergWitten}. 
This can be understood from the viewpoint of worldsheet open string theory 
in terms of commutation relations of target space coordinates 
\cite{ChuHo}. 
As all the fields living on D-branes are originated from 
degrees of freedom associated to the endpoints of open strings, 
the D-brane worldvolume theory exhibits 
the noncommutative nature of the endpoint coordinates. 
It is very interesting that the dynamical effect of $B$-field background 
(with vanishing field strength $H$) 
can be conveniently encoded in a geometric notion. 
It is tempting to conjecture that this success of 
geometrization of interactions admit further generalization. 

An immediate generalization considered in the literature
was the case of a nonvanishing field strength $H=dB$ of the $B$-field
\cite{Cornalba}. 
Since the Poisson limit of the commutation relation 
of target space coordinates is roughly given by 
the inverse of $B$, 
$H\neq 0$ suggests that $B$ is not a closed 2-form, 
or equivalently that $B^{-1}$ does not satisfy the Jacobi identity.
Naively this implies that the algebra of functions 
on the D-brane worldvolume becomes nonassociative, 
and it was shown \cite{Cornalba} 
that the naive nonassociative algebra is 
convenient for writing down some correlation functions. 
On the other hand, a proper canonical quantization would 
always define an associative algebra, 
and the correct associative algebra is at least equally good 
for expressing correlation functions \cite{Ho:2001qk}. 

Another direction for generalization is to consider 
the analogous problem in M theory. 
Type \IIA superstring theory can be interpreted 
as the 11 dimensional M theory with one spatial dimension 
compactified on a circle, 
and fundamental strings in \IIA superstring correspond to 
membranes in M theory, 
while $B$-field corresponds to $C$-field. 
The natural question is: 
what is the upgraded notion of noncommutative D-brane
in M theory? 

The canonical quantization of an open membrane ending on an M5-brane 
in a $C$-field background has been extensively studied in the literature
\cite{Berman,membraneC}.
(In particular the large $C$-field limit was studied in \cite{Berman}.)
The coordinates $X(\sigma)$ on the boundary of the open membrane, 
which is a closed string on M5-brane, 
satisfy a noncommutative relation. 
This algebraic structure is associated with the Poisson bracket, 
rather than the Nambu-Poisson bracket to be discussed below. 
A disadvantage of the study on canonical quantization is that 
there are constraints involved and the result depends on the choice of gauge. 
On the other hand the relevance of Nambu-Poisson bracket and 
volume-preserving diffeomorphism to M5-brane theory was 
already anticipated \cite{Matsuo} at that time.
A review of the M5-brane physics before the recent developments 
that will be focused on below can be found in \cite{Berman-review}.

In the study of D-branes in $B$-field background, 
there are other ways to discover D-brane noncommutativity 
besides quantizing target space coordinates of an open string. 
Alternatively one can compute 3-point correlation functions for 
generic vertex operators 
and realize that the effect of the $B$-field background 
is a universal exponential factor that coincides 
with the Moyal product of functions \cite{Schomerus}.
(There is no correction to 2-point correlation functions.)
Such a derivation can be most easily carried out 
in a low energy limit where higher oscillation modes on 
the string are ignored \cite{YinBigatti}. 
The analogous computation of correlation functions 
of open membranes was carried out in \cite{Ho:2007vk}
in a large $C$-field background. 
The result is again a universal exponential factor, 
but this time for 4-point correlation functions. 
For a 4-point scattering amplitude with 
external momenta $k_1, k_2, k_3, k_4=-k_1-k_2-k_3$, 
the exponential factor is 
\be
e^{\pm iC^{-1/2}\sqrt{k_1\cdot(k_2\times k_3)}}.
\ee
(There is no correction to 3-point correlation functions.)
The total amplitude is a sum over exponential factors of 
both signs in the exponent, giving
\be
\cos\left(C^{-1/2}\sqrt{k_1\cdot(k_2\times k_3)}\right)
\simeq 1 - \frac{1}{2C} k_1\cdot(k_2\times k_3) + \cdots. 
\ee
According to this formula, 
we expect that the M5-brane worldvolume action, 
which is a low energy description of open-membrane degrees of freedom, 
should be modified by interaction terms involving such factors
\be
\frac{1}{2C} \; \eps^{ijk} \; \del_i\otimes\del_j\otimes\del_k 
\label{Cdel3}
\ee
due to turning on a constant $C$-field background. 
Interestingly, 
this is precisely the Nambu-Poisson structure 
considered long time ago by Nambu \cite{Nambu}, 
and later formally defined by Takhtajan \cite{Takhtajan}, 
as a generalization of the Poisson structure in 
the Hamiltonian formulation. 
We will see below more precisely how the Nambu-Poisson structure 
dictates the gauge symmetry and interactions for 
an M5-brane in a large $C$-field background.
But let us conclude this section by giving the mathematical definition 
of Nambu-Poisson bracket. 

For an $n$-dimensional manifold ${\cal M}$, 
a tri-linear map $\{\cdot, \cdot, \cdot\}$ that maps 
3 functions on ${\cal M}$ to a single function on ${\cal M}$ 
is called Nambu-Poisson bracket if 
\footnote{
Apart from the Leibniz rule, 
this is the same definition for Lie 3-algebra 
\cite{Filippov,HHM}.
}
\begin{enumerate}
\item 
It is totally antisymmetrized 
\be
\{ f_1, f_2, f_3 \} = - \{ f_2, f_1, f_3 \} = - \{ f_1, f_3, f_2 \}. 
\ee
\item 
It satisfies the Leibniz rule
\be
\{ f_1, f_2, g_1 g_2 \} = \{ f_1, f_2, g_1 \} g_2 + g_1 \{ f_1, f_2, g_2 \}.
\ee
\item 
It satisfies the generalized Jacobi identity (fundamental identity)
\be
\{ f_1, f_2, \{ g_1, g_2, g_3 \} \} 
= \{ \{ f_1, f_2, g_1 \}, g_2, g_3 \} 
+ \{ g_1, \{ f_1, f_2, g_2 \}, g_3 \} 
+ \{ g_1, g_2, \{ f_1, f_2, g_3 \} \}.
\ee
\end{enumerate}
The most important theorem about Nambu-Poisson bracket 
is the decomposition theorem \cite{Jacobian}, 
which states that any Nambu-Poisson bracket 
can be locally expressed as 
\be
\{ f, g, h \} = \eps^{ijk} \; \del_i f \, \del_j g \, \del_k h 
\ee
for 3 coordinate $x^1, x^2, x^3$ out of the $n$ coordinates. 
From the viewpoint of constructing an M5-brane from 
infinitely many M2-branes via the BLG model, 
the decomposition theorem explains why 
there exist only M5-branes but not M8-branes nor other M-branes in M theory.
For a review of the mathematical properties of Nambu-Poisson bracket, 
see \cite{NPreview}.

\section{M5-brane in large $C$-field background}
\label{M5}

The worldvolume coordinates for an M5-brane 
in a large constant $C$-field background 
are naturally divided into two sets:
$\{x^{\mu} = x^1, x^2, x^3\}$ and $\{y^{\dmu} = y^{\dot{1}}, y^{\dot{2}}, y^{\dot{3}}\}$, 
so that $C_{\mu\nu\lam}$ and $C_{\dmu\dnu\dlam}$ are nonzero, 
but those components with mixed indices, 
$C_{\mu\nu\dlam}$ and $C_{\mu\dnu\dlam}$, vanish.
Although there is a symmetry between $x^{\mu}$ and $y^{\dmu}$ 
because $C$ is self-dual, 
the formulation given below is asymmetric. 

The most salient feature of M5-brane theory is that 
there is a self-dual 2-form gauge potential $b$, 
which is composed of 3 types of components 
$b_{\mu\nu}, b_{\dmu\dnu}$ and $b_{\mu\dmu}$. 
However, due to the self-duality condition, 
as we will see below, 
it is sufficient to explicitly keep only $b_{\dmu\dnu}$ and $b_{\mu\dmu}$, 
with $b_{\mu\nu}$ hidden in the formulation.
The hidden components $b_{\mu\nu}$ 
can be constructed on the way to solve equations of motion.

The field content of the M5-brane theory includes 
the transverse coordinates of the M5-brane $X^i$ $(i = 1, 2, \cdots, 5)$, 
the self-dual 2-form gauge potential $b$ 
and 6D chiral fermions $\Psi$, 
which can be conveniently organized as 
a single 11D Majorana spinor satisfying 
the 6D chirality condition
\be
\Gamma^7\Psi=-\Psi,
\ee
with the chiality matrix $\Gamma^7$ defined by
\be
\Gamma^{\mu\nu\rho}\Gamma^{\dot1\dot2\dot3}=\epsilon^{\mu\nu\rho}\Gamma^7.
\ee
(These are 11D $\gamma$-matrices.)

A 2-form gauge potential in 6D has $C^4_2 = 6$ polarizations. 
The self-duality condition reduces it to 3 independent components.
Together with the 5 scalars $X^i$, 
there are 8 bosonic degrees of freedom. 
An 11D Majorana fermion has 32 real components 
and the 6D chirality condition reduces it to 
16 independent fermionic degrees of freedom
as the superpartners of the 8 bosons.
(Two fermionic degrees of freedom is equivalent to one bosonic degree of freedom, 
since the Dirac equation is first order and Klein-Gordon equation is 2nd order.)
The field content of the M5-brane theory constitutes 
a tensor multiplet of the 6 dimensional ${\cal N}$ = (2, 0) supersymmetry.

\subsection{Action}

The action of a single M5-brane in a $C$-field background is 
derived from the Bagger-Lambert action \cite{BL}
with the Lie 3-algebra chosen to be the Nambu-Poisson algebra \cite{M51}. 
It is found to be \cite{M52}
\be
S = \frac{T_{M5}}{g^2}
\left( S_{\mbox{\em \small boson}} + S_{\mbox{\em \small ferm}} + S_{CS} \right), 
\label{M5S}
\ee
where $T_{M5}$ is the M5-brane tension and
\footnote{
$\Psi$ here was denoted by $\Psi'$ in \cite{M52}.
}
\begin{eqnarray}
S_{\mbox{\em \small boson}}
&=&\int d^3 x d^3 y \; \left[
-\frac{1}{2}({\cal D}_\mu X^i)^2
-\frac{1}{2}({\cal D}_{\dot\lambda}X^i)^2
-\frac{1}{4}{\cal H}_{\lambda\dot\mu\dot\nu}^2
-\frac{1}{12}{\cal H}_{\dot\mu\dot\nu\dot\rho}^2
\right.\nonumber\\&&\left.
-\frac{1}{2g^2}
-\frac{g^4}{4}\{X^{\dot\mu},X^i,X^j\}^2
-\frac{g^4}{12}\{X^i,X^j,X^k\}^2\right], 
\label{Sboson}
\\
S_{\mbox{\em \small ferm}}
&=&\int d^3 x d^3 y \; \left[
\frac{i}{2}\ol\Psi\Gamma^\mu {\cal D}_\mu\Psi
+\frac{i}{2}\ol\Psi\Gamma^{\dot\rho}{\cal D}_{\dot\rho}\Psi
\right.\nonumber\\&&\left.
+\frac{ig^2}{2}\ol\Psi\Gamma_{\dot\mu i}\{X^{\dot\mu},X^i,\Psi\}
-\frac{ig^2}{4}\ol\Psi\Gamma_{ij}\Gamma_{\dot1\dot2\dot3}\{X^i,X^j,\Psi\}
\right], 
\label{Sfermi0} 
\\
S_{CS}
&=&
\int d^3 x d^3 y \; 
\epsilon^{\mu\nu\lambda}\epsilon^{\dot\mu\dot\nu\dot\lambda}
\left[ -\frac{1}{2}
\partial_{\dot\mu}b_{\mu\dot\nu}\partial_\nu b_{\lambda\dot\lambda}
+\frac{g}{6}
\partial_{\dot\mu}b_{\nu\dot\nu}
\epsilon^{\dot\rho\dot\sigma\dot\tau}
\partial_{\dot\sigma}b_{\lambda\dot\rho}
(\partial_{\dot\lambda}b_{\mu\dot\tau}-\partial_{\dot\tau}b_{\mu\dot\lambda})
\right].
\label{CSt}
\end{eqnarray}

The covariant derivatives are defined by 
\ba
{\cal D}_\mu\Phi
&\equiv&\partial_\mu\Phi
-g\{b_{\mu\dot\nu},y^{\dot\nu},\Phi\},
\qquad
(\Phi = X^i, \Psi,)
\label{dmu}
\\
{\cal D}_{\dot\mu}\Phi
&\equiv&\frac{g^2}{2}\epsilon_{\dot\mu\dot\nu\dot\rho}
\{X^{\dot\nu},X^{\dot\rho},\Phi\}, 
\label{ddotmu} 
\ea
where
\be
X^{\dot\mu}(y)
\equiv\frac{y^{\dot\mu}}{g}
+
\frac{1}{2}\epsilon^{\dot\mu\dot\kappa\dot\lambda}
b_{\dot\kappa\dot\lambda}(y).
\ee

The field strengths are defined by
\begin{eqnarray}
{\cal H}_{\lambda\dot\mu\dot\nu}
&=&\epsilon_{\dot\mu\dot\nu\dot\lambda}{\cal D}_\lambda X^{\dot\lambda}
\nonumber\\
&=&H_{\lambda\dot\mu\dot\nu}
-g\epsilon^{\dot\sigma\dot\tau\dot\rho}
(\partial_{\dot\sigma}b_{\lambda\dot\tau})
\partial_{\dot\rho}b_{\dot\mu\dot\nu},\label{h12def}\\
{\cal H}_{\dot1\dot2\dot3}
&=&g^2\{X^{\dot1},X^{\dot2},X^{\dot3}\}-\frac{1}{g}
\nonumber\\
&=&H_{\dot1\dot2\dot3}
+\frac{g}{2}
(\partial_{\dot\mu}b^{\dot\mu}\partial_{\dot\nu}b^{\dot\nu}
-\partial_{\dot\mu}b^{\dot\nu}\partial_{\dot\nu}b^{\dot\mu})
+g^2\{b^{\dot1},b^{\dot2},b^{\dot3}\}, 
\label{h30def}
\end{eqnarray}
where $H$ is the linear part of the field strength
\begin{eqnarray}
H_{\lambda\dot\mu\dot\nu}
&=&
\partial_{\lambda}b_{\dot\mu\dot\nu}
-\partial_{\dot\mu}b_{\lambda\dot\nu}
+\partial_{\dot\nu}b_{\lambda\dot\mu},\\
H_{\dot\lambda\dot\mu\dot\nu}
&=&
\partial_{\dot\lambda}b_{\dot\mu\dot\nu}
+\partial_{\dot\mu}b_{\dot\nu\dot\lambda}
+\partial_{\dot\nu}b_{\dot\lambda\dot\mu}.
\end{eqnarray}

\subsection{Symmetries}

The M5-brane action (\ref{M5S}) respects 
the worldvolume translational symmetry, 
the global $SO(2,1)\times SO(3)$ rotation symmetry, 
the gauge symmetry of volume-preserving diffeomorphisms 
and the 6D ${\cal N}$ = (2, 0) supersymmetry.

\subsubsection{Gauge symmetry}

The gauge transformation laws are
\begin{eqnarray}
\delta_{\Lambda}\Phi
&=&g \kappa^{\dot\rho}\partial_{\dot\rho}\Phi, \label{gt1}
\qquad
(\Phi = X^i, \Psi,)
\\
\delta_{\Lambda}b_{\dot\kappa\dot\lambda}
&=&\partial_{\dot\kappa}\Lambda_{\dot\lambda}
-\partial_{\dot\lambda}\Lambda_{\dot\kappa}
+g\kappa^{\dot\rho}\partial_{\dot\rho} b_{\dot\kappa\dot\lambda},
\label{gt2}
\\
\delta_{\Lambda} b_{\lambda\dot\sigma}
&=&\partial_\lambda\Lambda_{\dot\sigma}
-\partial_{\dot\sigma}\Lambda_\lambda
+g\kappa^{\dot\tau}\partial_{\dot\tau}b_{\lambda\dot\sigma}
+g(\partial_{\dot\sigma}\kappa^{\dot\tau})b_{\lambda\dot\tau}, \label{gt4}
\end{eqnarray}
where
\be
\kappa^{\dot\lambda}\equiv
\epsilon^{\dot\lambda\dot\mu\dot\nu}\partial_{\dot\mu}
\Lambda_{\dot\nu}(x,y).
\ee
Eq.(\ref{gt2}) and (\ref{gt4}) can be more concisely expressed in terms of 
\ba
b^{\dmu} &\equiv& \eps^{\dmu\dnu\dlam} b_{\dnu\dlam}, \\
B_{\mu}{}^{\dmu} &\equiv& \eps^{\dmu\dnu\dlam}\del_{\dnu}b_{\mu\dlam}
\ea
as 
\ba
\d_{\Lambda} b^{\dmu} &=& \kappa^{\dmu} + g\kappa^{\dnu}\del_{\dnu} b^{\dmu}, 
\\
\d_{\Lambda} B_{\mu}{}^{\dmu} &=& 
\del_{\mu}\kappa^{\dmu} + g\kappa^{\dnu}\del_{\dnu}B_{\mu}{}^{\dmu}
- g(\del_{\dnu}\kappa^{\dmu})B_{\mu}{}^{\dnu}.
\ea

In terms of $B_{\mu}{}^{\dmu}$, 
the covariant derivative ${\cal D}_{\mu}$ acts as 
\be
{\cal D}_{\mu} \Phi = \del_{\mu} \Phi - gB_{\mu}{}^{\dmu}\del_{\dmu} \Phi.
\ee

While $b^{\dmu}$ determines $b_{\dnu\dlam}$ uniquely, 
$B_{\mu}{}^{\dmu}$ does not determine $b_{\mu\dnu}$ uniquely. 
Nevertheless, with the constraint 
\be
\del_{\dmu} B_{\mu}{}^{\dmu} = 0, 
\label{dB0}
\ee
$b_{\mu\dnu}$ can be determined by $B_{\mu}{}^{\dmu}$ 
up to a gauge transformation.
Therefore, the physical degrees of freedom represented by 
$b_{\dmu\dnu}$ and $b_{\mu\dnu}$
can be equivalently represented by $b^{\dmu}$ and $B_{\mu}{}^{\dmu}$.
In terms of $X^i, \Psi, b^{\dmu}$ and $B_{\mu}{}^{\dmu}$, 
all gauge transformations can be expressed solely in terms of $\kappa^{\dmu}$, 
without referring to $\Lambda_{\dmu}$ at all, 
as long as one keeps in mind the constraint 
\be
\del_{\dmu}\kappa^{\dmu} = 0
\ee
on the gauge transformation parameter $\kappa^{\dmu}$.

This gauge transformation can be interpreted as 
volume-preserving diffeomorphism
\be
\d y^{\dmu} = \kappa^{\dmu}, 
\qquad \mbox{with} \qquad
\del_{\dmu} \kappa^{\dmu} = 0.
\ee

\subsubsection{Supersymmetry}

The SUSY transformation laws are
\footnote{
$\eps$ here was denoted by $\eps'$ 
in \cite{M52}.
}
\begin{equation}
\delta^{(1)}_\chi \Psi=\chi,\quad
\delta^{(1)}_\chi X^i=\delta^{(1)}_\chi b_{\dot\mu\dot\nu}
=\delta^{(1)}_\chi b_{\mu\dot\nu}=0.
\label{chitr}
\end{equation}
and
\begin{eqnarray}
\delta^{(2)}_\eps X^i
&=&i\ol\epsilon\Gamma^i\Psi,
\label{dX}
\\
\delta^{(2)}_\eps \Psi
&=&{\cal D}_\mu X^i\Gamma^\mu\Gamma^i\epsilon
+{\cal D}_{\dot\mu}X^i\Gamma^{\dot\mu}\Gamma^i\epsilon
\nonumber\\&&
-\frac{1}{2}
{\cal H}_{\mu\dot\nu\dot\rho}
\Gamma^\mu\Gamma^{\dot\nu\dot\rho}\epsilon
-\frac{1}{g}\left(1+g{\cal H}_{\dot1\dot2\dot3}\right)
\Gamma_{\dot1\dot2\dot3}\epsilon
\nonumber\\&&
-\frac{g^2}{2}\{X^{\dot\mu},X^i,X^j\}
\Gamma^{\dot\mu}\Gamma^{ij}\epsilon
+\frac{g^2}{6}\{X^i,X^j,X^k\}
\Gamma^{ijk}\Gamma^{\dot1\dot2\dot3}\epsilon,
\label{dPsi}
\\
\delta^{(2)}_\eps b_{\dot\mu\dot\nu}
&=&-i(\ol\epsilon\Gamma_{\dot\mu\dot\nu}\Psi),
\label{db1}
\\
\delta^{(2)}_\eps b_{\mu\dot\nu}
&=&-i\left(1+g{\cal H}_{\dot1\dot2\dot3}\right)
\ol\epsilon\Gamma_\mu\Gamma_{\dot\nu}\Psi
+ig(\ol\epsilon\Gamma_\mu\Gamma_i\Gamma_{\dot1\dot2\dot3}\Psi)
\partial_{\dot\nu}X^i.
\label{db2}
\end{eqnarray}
Like $\Psi$, 
the SUSY transformation parameters can be 
conveniently denoted as an 11D Majorana spinor
satisfying the 6D chirality condition
\be
\Gamma^7\chi=\chi, \qquad
\Gamma^7\epsilon=\epsilon.
\ee

Both $\d^{(1)}$ and $\d^{(2)}$ are nonlinear SUSY transformations. 
But there is a linear combination 
\be
\d^{(1)}_\chi - g \d^{(2)}_\eps 
\qquad \mbox{with} \qquad
\chi = \Gamma^{\dot{1}\dot{2}\dot{3}}\eps
\ee
that defines a linear SUSY transformation,
a symmetry respected by the M5-brane vacuum state. 

In (\ref{dPsi}) and (\ref{db2}), we notice the appearance 
of the combination 
$({\cal H}_{\dot{1}\dot{2}\dot{3}} + g^{-1})$.
This should be taken as a hint that 
$1/g$ is the background $C$-field, 
being reminiscent of the combination $(dB + C)$ in the old M5-brane theory.
Similarly, in the action (\ref{Sboson}) 
there is a term $-1/(2g^2)$, 
which should be interpreted as the contribution 
of the background $C$-field. 
This is to be compared with the kinetic term $-(d B + C)^2/2$
of the gauge field in the old formulation,  
with the cross term $CdB$ ignored as a total derivative.
By comparing the M5-brane action with the noncommutative D4-brane action 
via double dimension reduction, 
one can verify the relation $C = 1/g$ more precisely \cite{M52}.
As each Nambu-Poisson bracket comes with a factor of $g$ 
in the M5-brane action, 
this is precisely the modification due to a $C$-field background 
expected in sec. \ref{NP} (see eq.(\ref{Cdel3})). 

The superalgebra of the SUSY transformations above has been investigated in \cite{Low}, 
where central charges of the superalgebra are derived, 
including the central charge for the soliton mentioned below in sec. \ref{soliton}.

\section{Comments}
\label{Comments}

\subsection{Seiberg-Witten map}

To relate the Abelian gauge symmetry in the old formulation of M5-brane
and the non-Abelian gauge symmetry in the new formulation, 
we use the notion of Seiberg-Witten map, 
which was originally proposed 
to connect gauge symmetries on commutative and noncommutative spaces
\cite{SeibergWitten}. 
Let the gauge transformation parameter be denoted $\lam$ and $\hat{\lam}$
for the Abelian and non-Abelian gauge transformations, respectively, 
the Seiberg-Witten map specifies the correspondence 
between two kinds of gauge transformations as follows
\be
\hat{\d}_{\hat{\lam}}\hat{\Phi}(\Phi) =
\hat{\Phi}(\Phi+\d_{\lam}\Phi) - \hat{\Phi}(\Phi),
\ee
where $\hat{\Phi}$ and $\Phi$ are corresponding fields 
in the two theories, and 
$\hat{\lam}$ is determined by $\lam$ and the gauge potential.
In this notation, all the fields in previous sections 
should wear hats. 
Corresponding fields in the old formulation come without hats.

The Seiberg-Witten map was determined to the first order in \cite{M52} as
\begin{eqnarray}
\hat{\Phi} &=& \Phi + g b^{\dot{\mu}} \partial_{\dot\mu}\Phi 
+ {\cal O}(g^2), \qquad (\Phi = X^i, \Psi) 
\\
\hat{b}^{\dot\mu}(b) &=& b^{\dot\mu} + 
\frac{g}{2} b^{\dot\nu}\partial_{\dot\nu} b^{\dot\mu} + 
\frac{g}{2} b^{\dot\mu} \partial_{\dot\nu} b^{\dot\nu} 
+ {\cal O}(g^2), 
\label{SWmb} 
\\
\hat{B}_{\mu}{}^{\dot\mu}(B, b) &=& B_{\mu}{}^{\dot\mu} + 
g b^{\dot\nu} \partial_{\dot\nu} B_{\mu}{}^{\dot\mu}  
- \frac{g}{2} b^{\dot\nu} \partial_{\mu}\partial_{\dot\nu} 
b^{\dot\mu} + \frac{g}{2} b^{\dot\mu} 
\partial_{\mu}\partial_{\dot\nu} b^{\dot\nu} 
+ g \partial_{\dot\nu} b^{\dot\nu} B_{\mu}{}^{\dot\mu} 
\nonumber 
\\
&&
- g \partial_{\dot\nu} b^{\dot\mu} B_{\mu}{}^{\dot\nu} 
- \frac{g}{2} \partial_{\dot\nu} b^{\dot\nu} 
\partial_{\mu} b^{\dot\mu} 
+ \frac{g}{2} \partial_{\dot\nu} b^{\dot\mu} 
\partial_{\mu} b^{\dot\nu} 
+ {\cal O}(g^2), 
\label{SWmB} \\ 
\hat{\kappa}^{\dot\mu}(\kappa, b) &=& \kappa^{\dot\mu} + 
\frac{g}{2} b^{\dot\nu}\partial_{\dot\nu} \kappa^{\dot\mu} 
+ \frac{g}{2} (\partial_{\dot\nu} b^{\dot\nu}) \kappa^{\dot\mu} 
- \frac{g}{2} (\partial_{\dot\nu} b^{\dot\mu}) \kappa^{\dot\nu} 
+ {\cal O}(g^2). \label{SWmk}
\end{eqnarray}
One can check that $\hat{\kappa}^{\dmu}$ satisfies 
the constraint $\del_{\dmu}\hat{\kappa}^{\dmu} = 0$ automatically
as long as $\kappa^{\dmu}$ satisfies 
the same constraint $\del_{\dmu}\kappa^{\dmu} = 0$.
Similarly, $\hat{B}_{\mu}{}^{\dmu}$ satisfies (\ref{dB0}) automatically.

The Seiberg-Witten map connecting commutative and noncommutative 
gauge theories are proven to exist to all orders. 
It remains to be checked whether the Seiberg-Witten map defined here 
can also be extended to all orders in $g$.

\subsection{Self-dual gauge theory}
\label{sd}

An interesting by-product of deriving M5-brane theory 
from the BLG model is the discovery of 
a new Lagrangian formulation of self-dual gauge theories. 
It is generally believed that there is no manifestly Lorentz-invariant 
Lagrangian formulation for self-dual gauge theories, 
unless auxiliary fields are introduced. 
(And the purpose of the auxiliary field is just to pick 
a special direction when it is gauge fixed.)
In the past, there exists non-manifestly Lorentz invariant Lagrangian formulation 
in which one out of the $D$ dimensions of the base space 
is chosen to play a special role in the formulation \cite{SD}. 
On the other hand, the new formulation divides the $d$ coordinates 
into two sets of coordinates of equal number.
(The spacetime dimensions of self-dual theories are always even.)
The M5-brane action at the quadratic level was derived in \cite{M51} 
and there the self-duality of the gauge field was explicitly checked.
The complete nonlinear M5-brane theory was given in \cite{M52}, 
including the self-dual gauge fields, 
but the self-duality of the gauge field strength was 
not explicitly checked until it was done in \cite{Pasti}, 
which also gave a good introduction to the basic ideas 
about self-dual gauge theories. 
Here we give a brief review of both the old and the new formulations 
at the quadratic level. 

In the old formulation we pick, say, the 0-th dimension 
(which could be either time-like or space-like)
to be special, 
and the action of the self-dual gauge field in 6 dimensions is 
\be
S = - \frac{1}{4!} \int d^6 x \; \left[
F_{\umu\unu\ulam}F^{\umu\unu\ulam} + 3 (F - \tilde{F})_{0ab} (F - \tilde{F})^{0ab}
\right], 
\label{S1}
\ee
where $\umu, \unu, \ulam = 0, \cdots, 5$, 
$a, b = 1, 2, \cdots, 5$, 
and $\tilde{F}$ denotes the dual of $F$
\be
\tilde{F}_{\umu\unu\ulam} = \frac{1}{3!} 
\eps_{\umu\unu\ulam\ua\ub\ug} F^{\ua\ub\ug}. 
\ee
In terms of the gauge potential $A_{\umu\unu}$, the field strength is 
\be
F_{\umu\unu\ulam} =
\del_{\umu}A_{\unu\ulam} + \del_{\unu}A_{\ulam\umu} + \del_{\ulam}A_{\umu\unu}, 
\ee
which is invariant under the gauge transformation 
\be
\d A_{\umu\unu} = \del_{\umu}\Lambda_{\unu} - \del_{\unu}\Lambda_{\umu}. 
\ee
The action (\ref{S1}) depends on $A_{0a}$ only through total derivative terms. 

The equations of motion derived from varying the action (\ref{S1}) 
with respect to $A_{ab}$ are 
\be
\eps^{abcde} \del_c ( F - \tilde{F} )_{0de} = 0. 
\ee
This implies that locally there exist functions $\Phi_a$ such that 
\be
( F - \tilde{F} ) _{0ab} = \del_a \Phi_b - \del_b \Phi_a. 
\ee
Now one can redefine $A_{0a}$ to absorb $\Phi_a$
by the replacement
\be
A_{0a} \rightarrow A_{0a} + \Phi_a, 
\ee
such that for the new gauge fields $A_{0a}, A_{ab}$, 
the self-duality conditions
\be
F_{0ab} = \tilde{F}_{0ab}
\ee
are satisfied.
(The condition $F_{abc} = \tilde{F}_{abc}$ is equivalent to this one.)

In the new formulation \cite{M52}, 
the base space coordinates are divided into two sets 
$\{ x^{\mu} \}$ and $\{ y^{\dmu} \}$. 
The action is 
\be
S = \frac{1}{4} \int d^6 x \;\left[
F_{\mu\dnu\dlam}(F - \tilde{F})^{\mu\dnu\dlam}
+ \frac{1}{3} F_{\dmu\dnu\dlam}(F - \tilde{F})^{\dmu\dnu\dlam} 
\right], 
\label{S2}
\ee
where $\mu, \nu, \lam = 1, 2, 3$ and $\dmu, \dnu, \dlam = \dot{1}, \dot{2}, \dot{3}$.
(One of the dimension is time-like, but it is unnecessary to know 
whether it belongs to $\{1, 2, 3\}$ or $\{\dot{1}, \dot{2}, \dot{3}\}$.)
The action above depends on $A_{ab}$ only through total derivatives. 
The equations of motion derived from varying (\ref{S2}) 
with respect to $A_{\dmu\nu}$ and $A_{\dmu\dnu}$ are 
\ba
&\del_{\dlam}(F-\tilde{F})^{\mu\dnu\dlam} = 0, 
\label{eq1}\\
&\del_{\mu} F^{\mu\dnu\dlam} + \del_{\dmu} F^{\dmu\dnu\dlam} = 0. 
\label{eq2}
\ea
The first equation implies that, locally, 
\be
(F-\tilde{F})_{\mu\dnu\dlam} = 
\frac{1}{2}\eps_{\dmu\dnu\dlam}\eps_{\mu\nu\lam}\del^{\dmu}\Phi^{\nu\lam}
\ee
for some $\Phi^{\mu\nu}$. 
Via a redefinition of $A_{\mu\nu}$ 
\be
A_{\mu\nu} \rightarrow A_{\mu\nu} + \Phi_{\mu\nu}, 
\ee
it becomes part of the self-duality conditions
\be
(F-\tilde{F})_{\mu\dnu\dlam} = 0.
\label{sd1}
\ee
This allows one to rewrite (\ref{eq2}) as a total derivative
\be
\del^{\dot{\mu}}(F_{\dmu\dnu\dlam} + 
\frac{1}{2}\eps_{\dmu\dnu\dlam}\eps_{\mu\nu\lam} \del^{\mu} A^{\nu\lam}) = 0, 
\ee
which can be easily solved as 
\be
F_{\dmu\dnu\dlam} + 
\frac{1}{2}\eps_{\dmu\dnu\dlam}\eps_{\mu\nu\lam} \del^{\mu} A^{\nu\lam}
= \eps_{\dmu\dnu\dlam} \Phi(x)
\label{eq10}
\ee
for some function $\Phi(x)$ 
which is independent of $y^{\dot{1}}, y^{\dot{2}}, y^{\dot{3}}$.
The function $\Phi(x)$ can thus be absorbed by a further redefinition of $A_{\mu\nu}$ 
\be
A_{\mu\nu} \rightarrow A_{\mu\nu} + \frac{1}{3} \eps_{\mu\nu\lam} \Phi_{\lam}(x),
\ee
where $\Phi_{\lam}(x)$ are chosen such that $\del_{\mu} \Phi^{\mu} = \Phi$.
Since $\Phi_{\lam}(x)$ can be chosen to be independent of $y^{\dmu}$, 
this further redefinition of $A_{\mu\nu}$ 
does not spoil the self-duality condition (\ref{sd1}) obtained earlier.
Thus (\ref{eq10}) becomes the remaining self-duality conditions
\be
F_{\dmu\dnu\dlam} = 
\frac{1}{6}\eps_{\dmu\dnu\dlam}\eps_{\mu\nu\lam} F^{\mu\nu\lam}.
\ee

For the self-dual gauge theory beyond the linearized equations, 
the readers are directed to \cite{M52} and \cite{Pasti}.

\subsection{Connection between M5-brane formulations}
\label{soliton}

A flat open membrane ending on an M5-brane is 
described as a string soliton in the M5-brane theory. 
The BPS solution was found in \cite{soliton-noB}, 
and later generalized to the case with $C$-field background in \cite{soliton-B}
for the old formulation of M5-branes. 
Its counterpart for the new formulation of M5-brane 
in $C$-field background was recently constructed in \cite{Furuuchi}
to the first order in $g$.
\footnote{
The complementary view of the string soliton 
from the multiple M2-branes' viewpoint 
was provided in \cite{Chu:2009iv}. 
}
There the Seiberg-Witten map discussed in a previous section 
was shown to correctly connect the two solutions in 
different formulations. 
This is so far the only explicit evidence supporting 
the equivalence of the two formulations of M5-branes 
in the limit of a large $C$-field background. 
More generally we expect a connection between 
the two formulations of M5-branes analogous 
to the connection between the DBI action 
and the noncommutative SYM action of D-branes. 
Perhaps the first step is to rewrite the old M5-brane action 
using the new formulation of self-dual gauge theory 
that decomposes the base space coordinates into 
two sets of 3 coordinates $x^{\mu}$ and $y^{\dmu}$.
Another direction to connecting the two formulations of M5-branes 
was suggested in \cite{BT1,BT2}, 
where some features of the new M5-brane theory are elucidated.

\subsection{Future study}
\label{Future}

In the above, we briefly reviewed the new M5-brane theory 
recently derived from the BLG model. 
This theory introduces a novel non-Abelian self-dual gauge theory 
defined by a Nambu-Poisson structure, 
and opens a new window to understanding 
the system of multiple M5-branes in M theory.
A lot of interesting questions remain to be answered. 
We conclude this article by giving an incomplete list
of topics for future study. 
\begin{enumerate}
\item
the quantization of Nambu-Poisson bracket.
\item
M5-brane action in {\em finite} $C$-field background.
\item
the action for the system of multiple M5-branes.
\item
the connection between the old and new formulations of M5-brane.
\item
the Seiberg-Witten map extended to higher orders in $g$.
\item
formulation of self-dual gauge theory in arbitrary dimensions
with generic separation of coordinates into two sets. 
\footnote{
In sec. \ref{sd} we mentioned decomposing $6$ dimensions 
as $1+5$ and as $3+3$.
It may be possible to find a formulation for $2+4$.
}
\end{enumerate}
While some of these are of a technical nature, 
others might be considered breakthroughs in theoretical physics.

\section*{Acknowledgment}

The authors thank David Berman, Chien-Ho Chen, Wei-Ming Chen, Chong-Sun Chu, 
Kazuyuki Furuuchi, Yosuke Imamura, 
Sangmin Lee, Yutaka Matsuo, 
Shotaro Shiba, Dmitri Sorokin and Tomohisa Takimi
for helpful discussions.
The author is supported in part by
the National Science Council,
Leung Center for Cosmology and Particle Astrophysics, 
and the National Center for Theoretical Sciences, Taiwan, R.O.C.

\end{document}